\documentclass[aps,pre,showpacs,twocolumn,groupedaddress]{revtex4-1}

\usepackage{graphicx,bbm,amsthm, amsmath, amssymb}
\usepackage{color}

\def\tr{\,{\rm tr}\,}
\def\ket#1{|#1\rangle}
\def\bra#1{\langle#1|}

\def\ave#1{\langle #1 \rangle}

\def\ii{{\rm i}}

\def\tit#1{}
\def\etal#1{ #1}

\def\Gr{\Gamma_{\rm R}}
\def\Gl{\Gamma_{\rm L}}

\begin{document}

\title{Solvable quantum nonequilibrium model exhibiting a phase transition and a matrix product representation}

\author{Marko \v Znidari\v c}
\affiliation{
Instituto de Ciencias F\' isicas, Universidad Nacional Aut\' onoma de M\' exico, C.~P.~62132, Cuernavaca, Morelos, Mexico, and\\
Department of Physics, Faculty of Mathematics and Physics, University of Ljubljana, Jadranska 19, SI-1000 Ljubljana, Slovenia}

\date{\today}

\begin{abstract}
We study a 1-dimensional XX chain under nonequilibrium driving and local dephasing described by the Lindblad master equation. The analytical solution for the nonequilibrium steady state found for particular parameters in {\em J.~Stat.~Mech., L05002 (2010)} is extended to arbitrary coupling constants, driving and homogeneous magnetic field. All one, two and three-point correlation functions are explicitly evaluated. It is shown that the nonequilibrium stationary state is not gaussian. Nevertheless, in the thermodynamic and weak-driving limit it is only weakly correlated and can be described by a matrix product operator ansatz with matrices of fixed dimension $4$. A nonequilibrium phase transition at zero dephasing is also discussed. It is suggested that the scaling of the relaxation time with the system size can serve as a signature of a nonequilibrium phase transition.
\end{abstract}

\pacs{05.30.-d, 03.65.Yz, 05.70.Ln, 75.10.Pq, 05.60.Gg}

\maketitle

\section{Introduction}
Interesting processes in physics are frequently associated with nonequilibrium situations. For instance, for a device to do some work an energy current must flow. The state of the system is therefore a nonequilibrium one. It is clearly desirable to understand nonequilibrium physics, unfortunately, as opposed to equilibrium physics, no general theory of nonequilibrium processes exists, for instance, for their stationary distribution. For equilibrium systems on the other hand the invariant equilibrium distribution is well known. The main difference between equilibrium and nonequilibrium systems is that in an equilibrium a principle of detailed balance holds, stating that the net flow of probability between any two states $x_1$ and $x_2$ is zero. That is, in equilibrium the probability to go from $x_1$ to $x_2$ is equal to the probability to go from $x_2$ to $x_1$. This simple rule greatly simplifies the analysis. In a stationary nonequilibrium situation on the other hand, only 
 the total probability flow out of a state $x_1$ has to be zero and not along each connection individually. In view of the lack of general theory it would be desirable to find some exact solutions for nonequilibrium situations, from which we could perhaps draw some generic rules. Particularly simple are nonequilibrium states that do not change in time, also called nonequilibrium stationary states (NESS).

In classical physics there are a number of exactly solvable nonequilibrium models, most notably various stochastic lattice gasses described by exclusion processes~\cite{exclusion,exDer,solvers}. The picture is quite different in quantum physics. There are hardly any analytically solvable models known, see though for instance an example of a single spin coupled to a bath in a star configuration~\cite{Breuer:04}. An exception of solvable models are those that are quadratic in the fermionic variables. One possibility for studying NESS is to take an infinite system in which an infinite sub-part serves as a bath. Such was the case in the doubly infinite XY chain studied in~\cite{XYinf}. The other approach is to write an effective master equation that describes the evolution of only the central system without baths. The simplest master equation is the Lindblad equation, which can be diagonalized if the superoperator (a generator of a markovian flow) is a quadratic function of fermi
 onic operators~\cite{3rd}, giving the exact solution for the NESS~\cite{Iztok08,bojan}.

In the present work we are going to find a solution for quantum NESS in a system whose Lindblad superoperator ${\cal L}$ (Eq.\ref{eq:Lin}) is not a quadratic function of fermions but is, nevertheless, simple enough to enable an explicit solution. Some parts of the solution have been presented in recent Refs.~\cite{JSTAT10,JPA10}, while the same model has been numerically studied in~\cite{NJP10}. The model we are going to study is the XX chain with a local dephasing at each site and coupled to nonequilibrium baths. A nonequilibrium XX model without dephasing was solved in~\cite{Karevski:09}, for a compact solution see also~\cite{JPA10}. Besides providing an explicit solution we also show that it is fundamentally different from those in quadratic systems. Namely, the Wick theorem does not hold and the NESS is not Gaussian. Another interesting aspect is that in a certain thermodynamic or weak driving limit the solution can be written as a matrix product operator (MPO) with matri
 ces of small fixed dimension $4$. This extends the MPO solution obtained for the NESS in a model without dephasing~\cite{JPA10}. The applicability of a matrix product ansatz has important implications for the properties of the system as well as for numerical methods that can be used to calculate NESS. It has been known for some time that the matrix product ansatz can describe ground states of certain 1-dimensional systems~\cite{MPold}, NESS of classical exclusion models~\cite{Derrida93,solvers}, and is a rather useful concept in quantum information and numerical methods used to simulate quantum systems~\cite{Murg}. A MPO ansatz of small fixed dimension can describe the time evolution of certain operators in integrable systems~\cite{pre:07,clark:10,new}. The MPO solution presented here extends the applicability of a MPO ansatz to NESSs in non-quadratic systems.

The final section deals with phase transitions in nonequilibrium stationary states. It is shown that the model exhibits a nonequilibrium phase transition at zero dephasing, going from the NESS that is Gaussian and displays a ballistic transport, to the NESS that is non-Gaussian, shows diffusive spin transport, and exhibits long-range correlations for nonzero dephasing. We show that the characteristic feature of the phase transition point is a faster closing of the gap of the superoperator with the system size than away from the phase transition. This seems to be a general property of nonequilibrium phase transitions, see recent studies in Refs.~\cite{Iztok08,PRL10}.

\section{The model}

The Hamiltonian of the XX spin chain in a homogeneous magnetic field is given by
\begin{equation}
H=\sum_{j=1}^{n-1} (\sigma_j^{\rm x} \sigma_{j+1}^{\rm x} +\sigma_j^{\rm y} \sigma_{j+1}^{\rm y})+B \sum_{i=1}^n \sigma_i^{\rm z},
\label{eq:H}  
\end{equation}
with standard Pauli matrices. An exact evolution of a central system, like the XX chain in our case, which is coupled to environment is in general complicated. However, if the environment correlation time is sufficiently small, i.e., so that it has no ``memory'', and the evolution has a property of the semigroup, then the reduced evolution of the central system can be described by the Lindblad equation~\cite{lindblad},  
\begin{equation}
\frac{{\rm d}}{{\rm d}t}{\rho}=\ii [ \rho,H ]+ {\cal L}^{\rm dis}(\rho)={\cal L}(\rho).
\label{eq:Lin}
\end{equation}
The dissipative linear operator ${\cal L}^{\rm dis}$ can be written in terms of Lindblad operators $L_k$,
\begin{equation}
{\cal L}^{\rm dis}(\rho)=\sum_k \left( [ L_k \rho,L_k^\dagger ]+[ L_k,\rho L_k^{\dagger} ] \right).
\end{equation}
The Lindblad master equation (\ref{eq:Lin}) can describe the most general completely positive trace preserving map that is a dynamical semigroup, i.e., a map that is a semigroup for a continuous time parameter. While the complete positivity must arguably be satisfied by any evolution, the semigroup property can be violated when the environment has a ``memory'' that causes a back-action on the central system. 

The model we study here has two different dissipative parts. One describee the action of two baths, inducing a nonequilibrium situation if they are different, while the other describes a local dephasing at each site, for instance, being due to the coupling of each site to unobservable degrees of freedom. The Lindblad dissipator is therefore a sum of two terms,
\begin{equation}
{\cal L}^{\rm dis}={\cal L}^{\rm bath}+{\cal L}^{\rm deph}.
\end{equation}
The dephasing part ${\cal L}^{\rm deph}=\sum_{j=1}^n{{\cal L}^{\rm deph}_j}$ is a sum of ${\cal L}^{\rm deph}_j$, each of which acts only on the $j$-th site and is described by a single Lindblad operator, 
\begin{equation}
L^{\rm deph}_j=\sqrt{\frac{\gamma}{2}}\sigma^{\rm z}_j.
\end{equation}
Sometimes it is useful to write a matrix representation of the dissipative superoperator ${\cal L}^{\rm deph}_j$. If we use a basis of Pauli matrices and we order them as $\{\sigma^{\rm x}_j,\sigma^{\rm y}_j,\sigma^{\rm z}_j,\mathbbm{1}_j\}$, we have a matrix representation
\begin{equation}
{\cal L}^{\rm deph}_j=
\left( \begin{array}{cccc}
-2\gamma & 0 & 0 & 0 \\
0 & -2\gamma & 0 & 0 \\
0 & 0 & 0 & 0 \\
0 & 0 & 0 & 0\\
\end{array} \right).
\end{equation}
Dephasing with strength $\gamma$ causes an exponential decay of the off-diagonal elements of a density matrix, if we write it in the diagonal basis of $\sigma^{\rm z}$. Using Jordan-Wigner transformation, such basis corresponds to the number basis of spinless fermions. Note that it is precisely the dephasing term which makes the superoperator non-quadratic, in fact quartic, in fermionic operators. Namely, the superoperator ${\cal L}^{\rm deph}$ is quadratic in the Lindblad operator $\propto \sigma^{\rm z}$ which is itself quadratic in fermionic operators.

For the dissipative bath part ${\cal L}^{\rm bath}$ we shall take the simplest possible Lindblad operators that are still able to describe a nonequilibrium situation. First, they are going to act locally only on the first and the last spin, and second, there will be only two operators at each end. Writing the dissipative part ${\cal L}^{\rm bath}={\cal L}^{\rm bath}_{\rm L}+{\cal L}^{\rm bath}_{\rm R}$ as a sum of a part acting only at the left end (site index $j=1$ and label ``L'') and a part acting only at the right end (site index $j=n$ and label ``R''), we have ${\cal L}^{\rm bath}_{\rm L,R}(\rho)=\sum_{k=1,2} \left( [ L_k^{\rm L,R} \rho,L_k^{{\rm L,R}\dagger} ]+[ L_k^{\rm L,R},\rho L_k^{{\rm L,R}\dagger} ] \right)$. The two Lindblad operators at the left end are
\begin{equation}
L^{\rm L}_1=\sqrt{\Gl(1-\mu+\bar{\mu})}\,\sigma^+_1,\qquad L^{\rm L}_2=\sqrt{\Gl(1+\mu-\bar{\mu})}\, \sigma^-_1,
\label{eq:Lbath}
\end{equation}
while on the right end we have
\begin{equation}
L^{\rm R}_1=\sqrt{\Gr(1+\mu+\bar{\mu})}\,\sigma^+_n,\qquad L^{\rm R}_2=\sqrt{\Gr(1-\mu-\bar{\mu})}\, \sigma^-_n,
\end{equation}
$\sigma^\pm_j=(\sigma^{\rm x}_j \pm {\rm i}\, \sigma^{\rm y}_j)/2$. The matrix representation of both superoperators is
\begin{eqnarray}
{\cal L}^{\rm bath}_{\rm L}&=&\Gl
\left( \begin{array}{cccc}
-2 & 0 & 0 & 0 \\
0 & -2 & 0 & 0 \\
0 & 0 & -4 & -4(\mu-\bar{\mu}) \\
0 & 0 & 0 & 0\\
\end{array} \right)\\ \nonumber \\ 
{\cal L}^{\rm bath}_{\rm R}&=&\Gr
\left( \begin{array}{cccc}
-2 & 0 & 0 & 0 \\
0 & -2 & 0 & 0 \\
0 & 0 & -4 & -4(-\mu-\bar{\mu}) \\
0 & 0 & 0 & 0\\
\end{array} \right).
\end{eqnarray}
Such simple local Lindbald operators involving $\sigma^+$ and $\sigma^-$ are often used when studying transport in spin chains, see for instance~\cite{Michel:03,Wichterich:07,Michel:08,JSTAT09,Steinigeweg:09}. Two parameters $\Gl$ and $\Gr$ play the role of a coupling strength, while $\mu$ and $\bar{\mu}$ determine the magnetization that the bath tries to impose on the chain ends. To see that this is indeed the case one can look for a stationary state of the bath ${\cal L}^{\rm bath}_{\rm L}$ dissipator only. That is, we look for a single-spin state $\tilde{\rho}$ that satisfies ${\cal L}^{\rm bath}_{\rm L}(\tilde{\rho})=0$. One easily finds that $\tilde{\rho} \sim \mathbbm{1}_1-(\mu-\bar{\mu})\sigma^{\rm z}_1$ and therefore $\tr{(\tilde{\rho}\sigma_1^{\rm z})}=-\mu+\bar{\mu}$. The chosen bath is therefore such that it tries to induce a local magnetization of size $-\mu+\bar{\mu}$ at the left end and $\mu+\bar{\mu}$ at the right end (these two parameters can be thought of as 
 magnetizations of an infinite bath to which the central system is coupled). Of course, the nonequilibrium stationary state of the whole master equation (\ref{eq:Lin}), which in addition includes a unitary part and a dephasing, will have a slightly different magnetization at the ends. One can also invert the problem and ask, is it possible to choose Lindblad operators that will target an arbitrary stationary state $\tilde{\rho}$ of ${\cal L}^{\rm bath}$, even on many qubits? The answer is positive with the explicit procedure given in~\cite{JSTAT09}.   

The goal of this paper is to analytically find the NESS of the master equation (\ref{eq:Lin}). This state, simply denoted by $\rho$ in the following, is an eigenstate of the Lindblad superoperator with an eigenvalue of $0$, ${\cal L}(\rho)=0$.

\section{The solution}

As shown in Ref~\cite{JSTAT10}, when $\bar{\mu}=0$ the solution, i.e. the NESS, can be sought in the form of a series in powers of the driving $\mu$. While a perturbative expansion is always possible, our solution is different. It is nonperturbative due to the special algebraic structure. We will discuss this point in more detail later. The ansatz for the NESS is
\begin{equation}
\rho=\frac{1}{2^n}\left( \mathbbm{1} + \mu R^{(1)}+\mu^2 R^{(2)}+\cdots \mu^r R^{(r)}+\cdots \right).
\label{eq:series}
\end{equation}
As we shall see, the term $\mu^r R^{(r)}$ is of the order $\mu^r$ and there is a closed set of equations that give $\mu^r R^{(r)}$. The solution can therefore be obtained term by term, without having to to deal with the whole set of exponentially many equations. We are first going to discuss the case with the zero bath magnetization offset, $\bar{\mu}=0$. The non-zero case will then be obtained as a simple modification of the solution for $\bar{\mu}=0$.

It was shown in Ref.~\cite{JSTAT10} that the solution $\rho$ is a sum of terms, where each is a product of operators $\sigma_j^{\rm z}$ and $j_k=2(\sigma_k^{\rm x} \sigma_{k+1}^{\rm y}-\sigma_k^{\rm y} \sigma_{k+1}^{\rm x})$ at different sites. $j_k$ is a spin current operator. This simple structure is a consequence of the fact, that all other operators, that could result in $j_k$ or $\sigma_j^{\rm z}$ when operated on by ${\cal L}$, are zero. The NESS can therefore be sought within the algebra of $\sigma_j^{\rm z}$ and $j_k$. Before actually going to the solution itself, let us discuss the role of a homogeneous magnetic field of strength $B$. 

The action of a single $B\, \sigma_j^{\rm z}$ term in the Hamiltonian is simple. The matrix of the superoperator ${\cal L}_j^{(\rm B)}$ has only two nonzero elements. They are ${\cal L}_j^{\rm (B)}(\sigma_j^{\rm x})= 2B\, \sigma_j^{\rm y}$, and ${\cal L}_j^{\rm (B)}(\sigma_j^{\rm y})=- 2B\,\sigma_j^{\rm x}$, while all other are zero, ${\cal L}_j^{\rm (B)}(\sigma_j^{\rm z,\mathbbm{1}})= 0$. Because the NESS is a sum of products of only $\sigma^{\rm z}_j$ and currents $j_k$, and because for homogeneous field we have $\sum_j {\cal L}_j^{\rm (B)}(j_k)= 0$ as well as $\sum_j {\cal L}_j^{\rm (B)}(j_k j_{k+1})= 0$, the NESS for zero field, $B=0$, is also an exact solution for arbitrary nonzero field $B \neq 0$. In short, the homogeneous magnetic field has no influence on the NESS.

\subsection{Hierarchy of connected correlations}
\label{sec:scaling}
Let us briefly argue why the NESS can be calculated term by term in the expansion over $\mu$, Eq.~(\ref{eq:series}), and why the set of equations for each term is closed. Without loss of generality we assume that $\bar{\mu}=0$. A key is a simple algebra generated by various superoperators in the master equation. The dephasing term acts as ${\cal L}^{\rm deph}(j_k)=-4\gamma j_k$ and ${\cal L}^{\rm deph}(\sigma_j^{\rm z})=\varnothing$, as well as ${\cal L}^{\rm deph}(\mathbbm{1}_j)=\varnothing$. The bath at the left end acts as ${\cal L}^{\rm bath}_{\rm L}(j_1)=-2j_1$, ${\cal L}^{\rm bath}_{\rm L}(\sigma_1^{\rm z})=-4\sigma_1^{\rm z}$, ${\cal L}^{\rm bath}_{\rm L}(\mathbbm{1}_1)=-4\mu \sigma_1^{\rm z}$. Similar expressions hold for the bath at the right end. The unitary part due to $H$ acts as ${\cal L}^{\rm H}_{k,k+1}(j_k)=8(\sigma_{k+1}^{\rm z}-\sigma_k^{\rm z})$, ${\cal L}^{\rm H}(\sigma_k^{\rm z})=(j_k-j_{k-1})$ and ${\cal L}^{\rm H}(\mathbbm{1}_k)=\varnothing$. Note that t
 here are no overlapping products of operators in $\rho$. The condition that there are no products of operators at the same site, i.e., either $\sigma_j^{\rm z} \sigma_j^{\rm z}$ or $j_k j_k$, ensures that the explicit normalization of $\rho$ by $1/2^n$ is not broken (\ref{eq:series}). Other terms not present due to hermiticity are $j_k \sigma_k^{\rm z}+\sigma_k^{\rm z} j_k \equiv \varnothing$ and $j_k \sigma_{k+1}^{\rm z}+\sigma_{k+1}^{\rm z} j_k \equiv \varnothing$; note however that $j_k j_{k+1}+j_{k+1}j_k \equiv -8(\sigma_k^{\rm x} \mathbbm{1}_{k+1}\sigma_{k+2}^{\rm x}+\sigma_k^{\rm y} \mathbbm{1}_{k+1}\sigma_{k+2}^{\rm y})$ are allowed. From the action of superoperators we note two things: (i) if $\tau$ is a sum of operators, each of which is a product of non-overlapping $\sigma_j^{\rm z}$ and $j_k$, then ${\cal L}(\tau)$ is again a sum of non-overlapping $\sigma_j^{\rm z}$ and $j_k$; and (ii) the number of operators $\sigma_j^{\rm z}$ and $j_k$ in each term is preserved
  by the action of ${\cal L}^{\rm H}$ and ${\cal L}^{\rm deph}$. The only superoperator that does not conserve the number of operators is that of the bath, which can create $\sigma^{\rm z}$ out of an identity. These two observations are crucial. Let us now write our solution $\rho$ as a series in $\mu$, Eq.~(\ref{eq:series}), where the $r$-th order term is a sum of all possible non-overlapping products, where each is a product of exactly $r$ operators $\sigma_j^{\rm z}$ or $j_k$ (here we always mean the number of non-identity operators). We put all (unknown) coefficients appearing in the term $\mu^r R^{(r)}$ in a set ${\cal S}^{(r)}$. For instance, the set $S^{(0)}$ has only one coefficient, namely a known normalization $1/2^n$ in front of $\mathbbm{1}$, the set $S^{(1)}$ consists of $n$ unknown coefficients in front of $\sigma_j^{\rm z}$ and $n-1$ unknown coefficients in front of $j_k$. The NESS must satisfy the equation ${\cal L}(\rho)=0$, therefore, the coefficient in fron
 t of each operator appearing in ${\cal L}(\rho)$ must be zero. If we look at the coefficient in front of an operator that is a product of $r$ non-identity operators, then this coefficient is a linear function of coefficients in the set $S^{(r)}$ and coefficients in the set $S^{(r-1)}$ (these come from the action of the bath ${\cal L}^{\rm bath}$). This structure enables us to solve for $\mu^r R^{(r)}$, i.e., coefficients in $S^{(r)}$, iteratively, starting with known $S^{(0)}$. Note that the only coefficient in $S^{(0)}$ is equal to $1/2^n \sim \mu^0$. Using this known $S^{(0)}$ we can now write a set of linear equations for coefficients in $S^{(1)}$, where the coefficient from $S^{(0)}$ will act as a source term, i.e., an inhomogeneous part of equations. Because when ${\cal L}^{\rm bath}$ makes a term of type $S^{(r+1)}$ from $S^{(r)}$, it always multiplies it by $\mu$, the source term will scale as $\mu^1$. Therefore, all coefficients in $S^{(1)}$ are proportional to $\mu^
 1$. Iteratively repeating the procedure we see that, (i) coefficients in $S^{(r)}$ scale as $\mu^r$, that is the terms in $\mu^r R^{(r)}$ indeed scale as $\mu^r$, and (ii) at each order we have to solve a closed set of equations for coefficients in $S^{(r)}$. These determine all $r$-point correlations in the NESS. Helping ourselves for the moment with the solution for the first three orders given in the following, we can say even more. If one writes equations for $r$-point connected correlations, instead of for non-connected ones, we can also predict how the source term at the $r$-th order scales with the number of spins $n$. Using equations from the previous order $S^{(r-1)}$ one can see that the source term scales as $\mu^r/n^r$. In the following we shall see that the spin current, the coefficient of which is denoted by $b$, scales in the same way. Therefore, the source terms scale as the $r$-th power of the current, $\sim b^r$. A consequence of this is that the largest co
 nnected term in $\mu^r R^{(r)}$ scales as $\sim b^r n$ and comes from the connected correlation of $r$ operators $\sigma_j^{\rm z}$. The largest non-connected term in $\mu^r R^{(r)}$ on the other hand scales as $\sim b^r n^r$ and comes from the product of $r$ operators $\sigma_j^{\rm z}$, see also the explicitly ansatz below.

We are now going to find an explicit form of the first three orders. For special values of parameters, the solution for the first two orders has been presented in~\cite{JSTAT10}.

\subsection{No magnetization offset, $\bar{\mu}=0$}

Let us first find the NESS for a bath with a zero offset of magnetization, $\bar{\mu}=0$. It is instructive to first find the equilibrium stationary state in the absence of driving, when $\mu=0$. In this case both baths (\ref{eq:Lbath}) act with $\sigma^+$ and $\sigma^-$ with equal probability, inducing no net magnetization. In fact, the equilibrium state is very simple and equal to the totally mixed state,
\begin{equation}
\rho_{\rm eq}=\frac{1}{2^n}\mathbbm{1}.
\end{equation}
This equilibrium state can therefore be thought of as an infinite temperature state.

\subsubsection{First two orders}

The Ansatz for the first two orders is the following,
\begin{equation}
\mu R^{(1)}=\mu A + \mu B,\qquad \mu A= \sum_{j=1}^n a_j \sigma^{\rm z}_j,\quad \mu B=\frac{b}{2}\sum_{k=1}^{n-1}j_k.
\label{eq:1st}
\end{equation}
The spin current operator is $j_k=2(\sigma_k^{\rm x} \sigma_{k+1}^{\rm y}-\sigma_k^{\rm y} \sigma_{k+1}^{\rm x})$. The 2nd order ansatz is
\begin{equation}
\mu^2 R^{(2)}=\frac{\mu^2}{2}\left(A B +B A \right)+\mu^2 C +\mu^2 D + \mu^2 F,
\end{equation}
\begin{eqnarray}
\mu^2 C&=&\sum_{j=1}^n \sum_{k=j+1}^n (C_{j,k}+a_j a_k) \sigma_j^{\rm z} \sigma_k^{\rm z},\nonumber \\
\mu^2 D&=&\sum_{j=1}^{n-2} \frac{d_j}{2} \left(\sum_{l=j+1}^{n-1}\sigma^{\rm z}_j j_l-\sum_{l=1}^{n-1-j}j_l \sigma^{\rm z}_{n+1-j} \right),\nonumber \\
\mu^2 F&=&\frac{f}{8}\sum_{k,l=1 \atop k\neq l}^{n-1} j_k j_l.
\end{eqnarray}
In the solution a specific factor will appear in all expressions for connected correlation functions. We denote it by the letter $t$,
\begin{equation}
t\equiv \frac{(\Gl+\Gr)(1+\Gl \Gr)+2(n-2)\gamma\Gl\Gr}{2\gamma \Gl \Gr}.
\label{eq:t}
\end{equation}
For large $n$ it scales as $t \sim n$, in fact, for $\Gl=\Gr=1$ it is equal to $n-2+2/\gamma$. It therefore plays the role of an effective system size.

The solution for the 1st and the 2nd order is found in exactly the same way as in Ref.~\cite{JSTAT10} for specific parameters. One writes a closed set of equations and solves it. We do not repeat the details of the derivation here, but just present the solution for general values of parameters. The 1st order terms are
\begin{equation}
b=-\mu\frac{2\Gl\Gr}{(\Gl+\Gr)(1+\Gl \Gr)+2(n-1)\gamma}=-\frac{\mu}{(t+1)\gamma},
\label{eq:b}
\end{equation}
and
\begin{eqnarray}
a_1&=&-\mu-\frac{b}{\Gl} \nonumber \\
a_2&=&a_1-b(\Gl+2\gamma) \nonumber \\
a_3&=&a_2-2\gamma b \nonumber \\
&\vdots & \nonumber\\
a_{n-1}&=& a_{n-2}-2\gamma b \nonumber \\
a_{n}&=& a_{n-1}-b(\Gr+2\gamma)=\mu +\frac{b}{\Gr}.
\label{eq:a}
\end{eqnarray}
Alternatively, we can express local magnetizations $a_j$ as 
\begin{eqnarray}
a_j&=&-\mu-b\,\, k^{(\rm L)}_j=\mu+b\,\, k^{(\rm R)}_j \\
k^{(\rm L)}_j&=&\{\frac{1}{\Gl},\frac{1+\Gl^2}{\Gl}+2\gamma,\ldots,\frac{1+\Gl^2}{\Gl}+2(n-1)\gamma+\Gr\}, \nonumber \\
k^{(\rm R)}_j&=&\{\frac{1+\Gr^2}{\Gr}+2(n-1)\gamma+\Gl,\ldots,\frac{1+\Gr^2}{\Gr}+2\gamma,\frac{1}{\Gr}\}. \nonumber \\
\label{eq:kj}
\end{eqnarray}
Away from the boundary we have $k_{j+1}^{\rm (L)}-k_{j}^{\rm (L)}=2\gamma$, while $k_{j-1}^{\rm (R)}-k_{j}^{\rm (R)}=2\gamma$. The expectation value of the magnetization at site $j$ is just $\ave{\sigma_j^{\rm z}}=a_j$ and of the spin current $\ave{j_k}=2b$. Several interesting observations can be made. Our one-spin Lindblad bath is constructed in such a way that it targets magnetization $\mp \mu$ on the 1st and the last site. Because there are other terms besides the bath in the Lindblad equation the actual magnetization in the NESS at the boundaries is slightly different. Namely, we have $a_1=-\mu-b/\Gl$ and $a_n=\mu+b/\Gr$. Expectedly, one can see that if the coupling strength $\Gl$ (or $\Gr$) is very small the difference from the targeted magnetization is large. Second, the magnetizaton difference between two neighboring sites is $2\gamma b$ in the bulk, while it is $b(2\gamma+\Gl)$ at the left end and $b(2\gamma+\Gr)$ at the right end. The magnetization profile is theref
 ore linear, apart from two sites at both ends, where a contact resistance due to $\Gl$ and $\Gr$ is felt. One could in fact generalize our solution to allow for an inhomogeneous dephasing $\gamma_j$ at each site. The difference in the first order solution (\ref{eq:a}) would be having terms $\gamma_j+\gamma_{j+1}$ instead of $2\gamma$. Using $\gamma_1=\gamma_n=0$ and having $\Gl=\Gr=\gamma$ one could therefore achieve a perfectly linear profile, without any contact resistance jumps at the boundaries. However, higher order terms in $\mu$ are more complicated in this case (although of the same form) and we do not discuss inhomogeneous dephasing in the present work. 

The expectation value of the current $2b$ is maximal for intermediate couplings $\Gamma_{\rm L,R}$. Keeping the system size and the dephasing strength $\gamma$ fixed, we can see that $b$ has a maximum at $\Gl=\Gr=1$. It is equal to $|b_{\rm max}|=\mu/(2+(n-1)\gamma)$. If the couplings are smaller the current is smaller because the NESS is only weakly nonequilibrium due to the weak coupling. On the other hand, if $\Gamma$s are large the current is again smaller because the hamiltonian part $H$, which transports magnetization, is less important compared to baths. The dependence of the maximal current on $\gamma$ is trivial, the smaller the dephasing the larger is the current. Another observation is that the magnetization difference between the two ends is $a_n-a_1=-b(\Gl+\Gr+2\gamma(n-1))$. The transport coefficient $\kappa$, defined via $j=-\kappa \frac{a_n-a_1}{n}$, is $\kappa=2n/(\Gl+\Gr+2(n-1)\gamma)$. As long as the dephasing is nonzero it asymptotically behaves as $\kappa
  \sim 1/\gamma$ and is independent of the system size $n$. If $n \gamma$ is on the other hand smaller than $\Gl,\Gr$, for instance if the dephasing is zero, then in the limit of weak coupling, $\Gl,\Gr \to 0$, $\kappa$ diverges.

The expressions for the 2nd order terms are rather simple,
\begin{equation}
f=b^2 (1+1/t)
\end{equation}
\begin{equation}
d_{i,j}=\frac{b^2}{t}
\left\{ \begin{array}{ccr}
-k^{(\rm L)}_i &;& \hbox{$j>i$}\\
k^{(\rm R)}_i &;& \hbox{$j<i$}\\
\end{array} \right. .
\end{equation}
\begin{equation}
C_{i,j}=-\frac{b^2}{t}\left( k^{(\rm L)}_i k^{(\rm R)}_j+ (1+t)\, \delta_{i+1,j}\right). 
\label{eq:C}
\end{equation}
The connected correlation function of magnetization, $\ave{\sigma_i^{\rm z} \sigma_j^{\rm z}}_{\rm c}=\ave{\sigma_i^{\rm z} \sigma_j^{\rm z}}-\ave{\sigma_i^{\rm z}}\ave{\sigma_j^{\rm z}}$ is for $i=j$ equal to $1-a_i^2$, while it is equal to $\ave{\sigma_i^{\rm z} \sigma_j^{\rm z}}_{\rm c}=C_{i,j}$ for nondiagonal $i \neq j$. For large system size $n$ the connected correlation function $C_{i,j}$ achieves its maximal value at $i\approx j \approx n/2$, when a product of $k^{(\rm L)}_i k^{(\rm R)}_j$ is the largest. We therefore have ${\rm max}(C_{i,j}) \sim (\gamma n b)^2/t$. The maximum is again achieved at $\Gl=\Gr=1$. The dependence of the maximal $z-z$ connected correlation on $\gamma$ is though the opposite of that of the current. Here the correlations monotonously increase with increasing $\gamma$.

The 3rd order terms are obtained in an analogous way. The calculation is a little tedious, nevertheless, we managed to obtained closed expressions. Detailed results can be found in the Appendix~\ref{app:A}. At this point we just list the dominant term for large $n$. It is a three-point connected correlation of $\sigma^{\rm z}$. Away from boundaries and for non-neighboring indices, where the Kronecker-delta terms are zero, the general expression (\ref{eq:zzz}) simplifies. For instance, for $\gamma=\Gl=\Gr=1$ and $i<j<k$ it is just
\begin{equation}
\ave{\sigma_i^{\rm z} \sigma_j^{\rm z}\sigma_k^{\rm z}}_{\rm c} = \frac{2(2\mu)^3}{n(n-1)}x(1-2y)(1-z),
\end{equation}
where we have introduced rescaled position variables $x=\frac{i}{n+1}, y=\frac{j}{n+1}, z=\frac{k}{n+1}$. The two-point function is for the same parameter values and $i<j$
\begin{equation}
\ave{\sigma_i^{\rm z} \sigma_j^{\rm z}}_{\rm c} = -\frac{(2\mu)^2}{n}x(1-y).
\label{eq:plat}
\end{equation}
Scaling of the first two connected correlations on $\mu$ and $n$ therefore follows a general rule: the $r$-point connected correlation function is $\sim \mu^r/n^{r-1}$ in accordance with the general discussion in the subsection~\ref{sec:scaling}. The form of these two correlation functions is the same as in some classical exclusion processes~\cite{exDer}, for a recent work on an analogous quantum master equation exhibiting some features similar to our model see~\cite{temme:09}. Whether this is only a consequence of the same hydrodynamic limit or whether there perhaps exists an exact mapping from the quantum model to a classical one is unknown at present.

\subsection{Nonzero offset, $\bar{\mu}\neq 0$}
Let us now go to the case where there is an offset of magnetization in the baths. For nonzero $\bar{\mu}$ the equilibrium state, when $\mu=0$, is again simple, although not trivial $\sim \mathbbm{1}$ as in the case of zero $\bar{\mu}$. It is a product state with nonzero magnetization
\begin{equation}
\rho_{\rm eq}=\frac{1}{2^n}\prod_{j=1}^n (\mathbbm{1}_j+\bar{\mu}\sigma_j^{\rm z}).
\end{equation}
It is therefore a state with all connected correlations being zero (trivially Gaussian state) apart from nonzero magnetization. This equilibrium state can be equated to the grandcanonical state $\rho_{\rm gc} \sim \exp{(-\beta(H- \chi \Sigma^{\rm z}))}$, with $\Sigma^{\rm z}=\sum_j \sigma_j^{\rm z}$, having an infinite temperature $\beta=0$ and a finite chemical potential times the inverse temperature, $\beta \,\chi=\frac{1}{2} \ln{((1+\bar{\mu})/(1-\bar{\mu}))}$. One should note though that integrable systems, such as our XX chain, do not thermalize for generic local Lindblad baths~\cite{PRE10}.

Out of equilibrium, when $\mu \neq 0$, the solution is actually very similar to the one with zero $\bar{\mu}$. Namely, the only thing that changes is the expression for magnetization, $a_j$, while all other $n$-point connected correlations are the same as in the case of no offset, $\bar{\mu}=0$. For nonzero offset the solution is therefore
\begin{equation}
a_j=\bar{\mu}-\mu-b\,\, k^{(\rm L)}_j,
\end{equation}
with the same $k^{(\rm L)}_j$ (\ref{eq:kj}) as before. All other terms (\ref{eq:b},\ref{eq:C},\ref{eq:zzz},\ref{eq:3rdX},\ref{eq:3rdY}) are the same.

\section{The NESS is non-gaussian}
In this section we are going to show that the NESS obtained above is not Gaussian, i.e., that the Wick theorem does not hold. Because three-point connected correlation functions are nonzero, see e.g. eq.~(\ref{eq:zzz}), it is obvious that the Wick theorem does not apply in spin variables. However, it is not clear that it does not apply in the spinless fermion picture. 

A system of spin-$1/2$ particles can be mapped to spinless fermions using the Jordan-Wigner transformation. Denoting by $c_j$ and $c_j^\dagger$ canonical fermionic annihilation and creation operators, satisfying anticommutators $\{c_j,c_k\}=0$, $\{c_j^\dagger,c_k^\dagger\}=0$, $\{c_j,c_k^\dagger\}=\delta_{j,k}\mathbbm{1}$, the transformation is given by the mapping
\begin{eqnarray}
c_j&=& -(\sigma_1^{\rm z}\cdots \sigma_{j-1}^{\rm z})\sigma_j^+ \nonumber \\
c_j^\dagger&=& -(\sigma_1^{\rm z}\cdots \sigma_{j-1}^{\rm z})\sigma_j^-,
\end{eqnarray}
or its inverse
\begin{eqnarray}
\sigma_j^{\rm x}&=& -(\sigma_1^{\rm z}\cdots \sigma_{j-1}^{\rm z})(c_j+c_j^\dagger) \nonumber \\
\sigma_j^{\rm y}&=& -\ii (\sigma_1^{\rm z}\cdots \sigma_{j-1}^{\rm z})(c_j-c_j^\dagger) \nonumber \\
\sigma_j^{\rm z}&=& c_j c_j^\dagger-c_j^\dagger c_j=\mathbbm{1}-2n_j,
\end{eqnarray}
where we denote by $n_j=c_j^\dagger c_j$ a number (density) operator at site $j$. Denoting by $Z_j^{(r)}=\sigma_j^{\rm z}\cdots \sigma_{j+r-1}^{\rm z}$ a string of $r$ consecutive $\sigma^{\rm z}$, and introducing an energy-density-like operator $H^{(r+1)}_j \equiv \sigma_j^{\rm x}Z_{j+1}^{(r-1)}\sigma_{j+r}^{\rm x}+\sigma_j^{\rm y}Z_{j+1}^{(r-1)} \sigma_{j+r}^{\rm y}$ and a current-like operator $B^{(r+1)}_j \equiv \sigma_j^{\rm x}Z_{j+1}^{(r-1)}\sigma_{j+r}^{\rm y}-\sigma_j^{\rm y}Z_{j+1}^{(r-1)} \sigma_{j+r}^{\rm x}$, we have
\begin{eqnarray}
H^{(r+1)}_j &=& 2(c_j^\dagger c_{j+r}-c_j c_{j+r}^\dagger) \nonumber \\
B^{(r+1)}_j &=& 2\ii (c_j^\dagger c_{j+r}+c_j c_{j+r}^\dagger) \nonumber \\
H^{-(r+1)}_j &=& 2(c_j^\dagger c_{j+r}^\dagger-c_j c_{j+r}) \nonumber \\
B^{-(r+1)}_j &=& 2\ii (c_j^\dagger c_{j+r}^\dagger+c_j c_{j+r}),
\end{eqnarray}
where $H^{-(r+1)}_j \equiv \sigma_j^{\rm x}Z_{j+1}^{(r-1)}\sigma_{j+r}^{\rm x}-\sigma_j^{\rm y}Z_{j+1}^{(r-1)} \sigma_{j+r}^{\rm y}$ and $B^{-(r+1)}_j \equiv \sigma_j^{\rm x}Z_{j+1}^{(r-1)}\sigma_{j+r}^{\rm y}+\sigma_j^{\rm y}Z_{j+1}^{(r-1)} \sigma_{j+r}^{\rm x}$. From our solution for the NESS (see also comments in ref.~(\cite{JSTAT10})) we can see that expectations of all $H^{(r)}_j,B^{(r+1)}_j,H^{-(r)}_j$ and $ B^{-(r)}_j$, apart from $B^{(2)}_j$, are zero. The only nonzero two-point fermionic expectations are therefore $\ave{c_j^\dagger c_j}=(1-a_j)/2$ and $2\ii\ave{c_j^\dagger c_{j+1}+c_j c_{j+1}^\dagger}=b$. An important point is that only on-site or nearest-neighbor two-point fermionic correlations are nonzero. If the NESS would be gaussian and the Wick theorem would hold all connected correlations beyond nearest neighbor would have to be zero. Because this is not the case the NESS is clearly not gaussian. For instance, rewriting the connected $z-z$ correlation as $C_{
 i,j}=\ave{\sigma_i^{\rm z}\sigma_j^{\rm z}}-\ave{\sigma_i^{\rm z}}\ave{\sigma_j^{\rm z}}=4(\ave{c_i^\dagger c_i \, c_j^\dagger c_j}-\ave{c_i^\dagger c_i}\ave{c_j^\dagger c_j})$, and using the Wick theorem, we would have $C_{i,j}=-\ave{c_i^\dagger c_j}\ave{c_i c_j^\dagger}$. The last expression is nonzero only if $j=i+1$, whereas on the other hand we have long-range correlations with all $C_{i,j}$ being nonzero if $\gamma \neq 0$. The Wick theorem therefore does not hold if $\gamma \neq 0$. The NESS is non-Gaussian and presents an interesting new solvable model. Whether it is nevertheless equivalent to some existing solvable model is at present unknown.

\subsection{Matrix product operator ansatz}
Even-though the NESS is non-Gaussian, it is still, in a sense, weakly correlated, meaning that it can be represented in terms of a matrix product operator (MPO) form
\begin{equation}
\rho=\frac{1}{2^n}\sum_{\alpha_j} \bra{1} A_1^{(\alpha_1)} A_2^{(\alpha_2)} \cdots A_n^{(\alpha_n)}\ket{1}\,\, \sigma_1^{\alpha_1} \sigma_2^{\alpha_2} \cdots \sigma_n^{\alpha_n},
\label{eq:MPOan}
\end{equation}
with matrices of small size. Note that in an MPO formulation a density matrix is treated as an element (a pure state) of $4^n$ dimensional Hilbert space of operators. For $\gamma=0$ the matrices $A_j^\alpha$ of finite size $D=4$ that is independent of the system length suffice~\cite{JPA10}. For nonzero dephasing an exact representation with matrices of fixed size is not possible. We have numerical indications though that the Schmidt rank, i.e., a number of nonzero Schmidt coefficients, for a bipartite cut after the first $m$ spins is $4m$. We have verified this conjecture by numerically computing an exact NESS for systems of up-to $n=10$ spins. Because the Schmidt rank is equal to the necessary dimension of matrices in the MPO ansatz, the exact representation of the NESS requires an MPO of dimension $D=2n$.   

Looking at the series solution for NESS (\ref{eq:series}) we can see that the largest connected term in the $r$-th order $\mu^r R^{(r)}$ (i.e., $r$-point connected correlation function) scales for large $n$ as $\sim b^r\,n=1/n^{r-1}$ and comes from the connected correlation of $r$ $\sigma^{\rm z}$s. Therefore, in the thermodynamic limit of large $n$ one could neglect all higher order connected correlations, keeping in the NESS only the leading terms that scale with $n$ as $\sim 1$ and $\sim 1/n$. These are magnetization $\sigma_j^{\rm z}$, spin current $j_k$ and $z-z$ connected correlations, as well as their products decaying no faster than $\sim 1/n$ that appear in higher order non-connected correlations. Because all these terms are already present in an exact MPO solution for $\gamma=0$~\cite{JPA10} having $D=4$, we can expect that in the limit of a large system one might be able to approximately describe the NESS with an MPO of dimension smaller than $2n$. An MPO with dime
 nsion $D=4$ that correctly describes all terms larger than or equal to $\sim {\cal O}(1/n)$ in the NESS is a simple extension of the solution for $\gamma=0$~\cite{JPA10}. Four matrices at each site have to have the following form, 
\begin{eqnarray}
A^{(z)}_i &=&
\left( \begin{array}{cccc}
a_i & c_i & 0 & 0 \\
s_i & a_i & 0 & 0 \\
0 & 0 & 0 & 0 \\
0 & 0 & 0 & 0\\
\end{array} \right), \quad 
A^{(\mathbbm{1})}_i =
\left( \begin{array}{cccc}
1 & 0 & 0 & 0 \\
0 & 1 & 0 & 0 \\
0 & 0 & 0 & 0 \\
0 & 0 & 0 & 0\\
\end{array} \right),\nonumber \\
A^{(\rm x)}_j &=&(\cos{\frac{\pi}{2}j}-\sin{\frac{\pi}{2}j})\,P, A^{(\rm y)}_j=-(\cos{\frac{\pi}{2}j}+\sin{\frac{\pi}{2}j})\,R \nonumber \\
P&=&
\left( \begin{array}{cccc}
0 & 0 & 0 & -b \\
0 & 0 & 0 & 0 \\
1 & 0 & 0 & 0 \\
0 & 0 & 0 & 0\\
\end{array} \right), \quad 
R=
\left( \begin{array}{cccc}
0 & 0 & -b & 0 \\
0 & 0 & 0 & 0 \\
0 & 0 & 0 & 0 \\
1 & 0 & 0 & 0\\
\end{array} \right),
\label{eq:MPO} 
\end{eqnarray}
where $c_i=\frac{b}{\sqrt{t}} k^{(\rm L)}_i$ and $s_i=-\frac{b}{\sqrt{t}} k^{(\rm R)}_i$. Parameters $a_i$ and $b$ are given in Eqs.(\ref{eq:a},\ref{eq:b}). Matrices $A^{(\rm x)}_i$ and $A^{(\rm y)}_i$ are periodic with period 4, $A^{(\rm x)}_{i+4}=A^{(\rm x)}_i$, and can be concisely written as $A^{(x)}=(-P,-P,P,P,-P,\ldots)$, and $A^{(y)}=(-R,R,R,-R,-R,\ldots)$, if one writes them as a vector. 

Whether the above MPO with $D=4$ actually suffices to describe the NESS in the thermodynamic limit depends on the scaling of the next largest Schmidt coefficient $\lambda_5$ (in this subsection we denote by $\lambda_j, j=1,\ldots$ the Schmidt coefficients for a symmetric bipartition at $n/2$ spins; $\sum_j \lambda_j^2=1$). If $D=4$ is to suffice, $\lambda_5$ must decay more rapidly than $\lambda_4$. Because the scaling of Schmidt coefficients is not simply related to the scaling of expansion coefficients of the NESS we used numerical simulation to study the scaling of $\lambda_j$. We have calculated the NESS using a time dependent density matrix renormalization group method~\cite{tdmrg}~\cite{NJP10} (tDMRG) for systems with up-to $n=128$ spins using an MPO ansatz with a fixed small matrix dimension $D=10$. In all results that follow we have fixed $\Gl=\Gr=\gamma=1$ and $\bar{\mu}=0$. As an independent check of our asymptotic MPO (\ref{eq:MPO}) solution, we also compared the f
 our largest Schmidt coefficients obtained from the tDMRG with those obtained from the $D=4$ MPO (\ref{eq:MPO}). Having an explicit representation of matrices (\ref{eq:MPO}) it is easy to calculate the Schmidt coefficients.

\begin{figure}[!h]
\centerline{\includegraphics[width=3.2in]{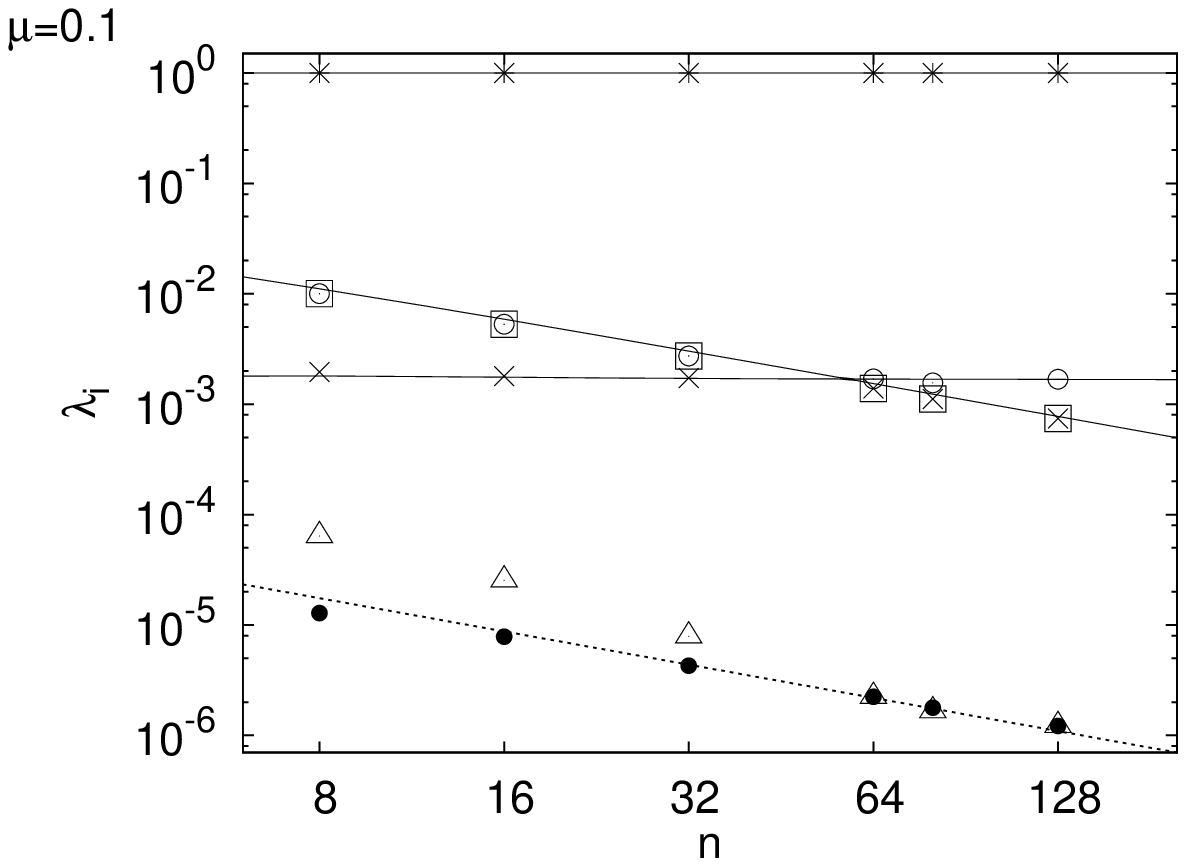}}
\centerline{\includegraphics[width=3.2in]{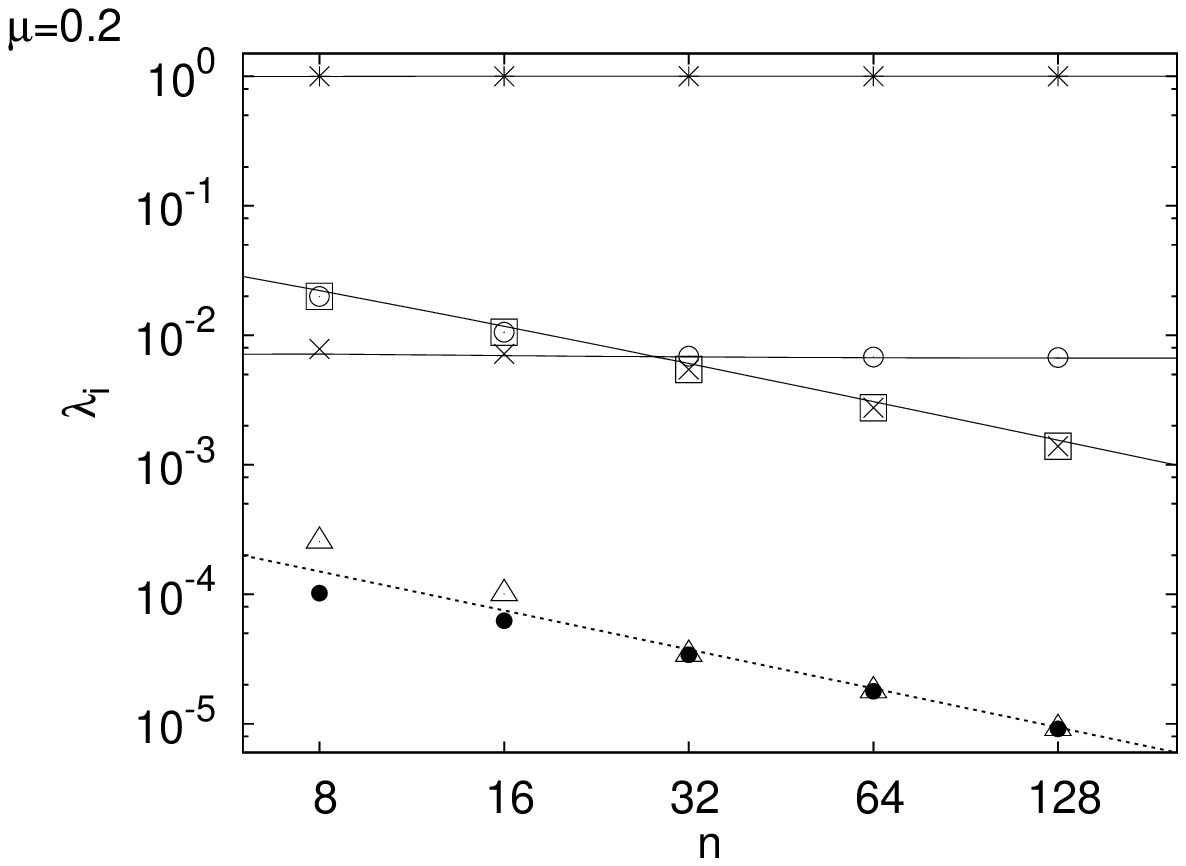}}
\caption{Scaling of the 6 largest Schmidt coefficients $\lambda_i$ of the NESS with the system size. Top figure is for $\mu=0.1$, bottom for $\mu=0.2$. Symbols are tDMRG results for $\lambda_1, \lambda_2, \lambda_3, \lambda_4, \lambda_5$ and $\lambda_6$ (stars, open circles, squares, crosses, triangles and full circles, respectively); full lines are analytic Schmidt coefficients from MPO with $D=4$ (\ref{eq:MPO}) while the dotted line indicates asymptotic $\sim 1/n$ scaling of $\lambda_5$. We use a symmetric bipartite cut after $n/2$ spins, $\Gl=\Gr=\gamma=1$, $\bar{\mu}=0$.}
\label{fig:spekter}
\end{figure}
The results for $\lambda_{1,\ldots,6}(n)$ are shown in Fig.~\ref{fig:spekter}. Several interesting things can be observed. First, the largest four Schmidt coefficients obtained by the tDMRG agree with the analytical calculation from Eq.(\ref{eq:MPO}). Interestingly, around $n\approx 60$ for $\mu=0.1$ and around $n\approx 30$ for $\mu=0.2$ an avoided crossing occurs between $\lambda_{2,3}$ and $\lambda_4$. A consequence of this is that for large $n$ we have scaling $\lambda_4 \sim 1/n$, while we have $\lambda_4 \approx n^0$ for small $n$. From the data for different $\mu$ we can infer that the crossing happens when $\mu \, n = \nu_{\rm c}$, with $\nu_{\rm c}\approx 6$. Similar avoided crossing at the same value of $n$ occurs also between $\lambda_5$ and $\lambda_6$. For $\mu\, n < \nu_{\rm c}$ $\lambda_5$ decays faster than $1/n$ while it decays as $\lambda_5 \sim 1/n$ for $\mu\, n > \nu_{\rm c}$. We therefore see that if we want $\lambda_5$ to decay with $n$ faster than $\lambda_4$, in other words, for $D=4$ MPO (\ref{eq:MPO}) to really give the leading term solution, one must have $\mu\, n < \nu_{\rm c}$ when going to the thermodynamic limit $n \to \infty$.

Another possibility for the $D=4$ solution to give the leading order would be a limit of small driving, $\mu \to 0$. If $\lambda_5$ decays with $\mu$ faster than $\lambda_4$ this would still be enough. We therefore also looked at the scaling of $\lambda_i$ with $\mu$. 
\begin{figure}[!h]
\centerline{\includegraphics[width=1.7in]{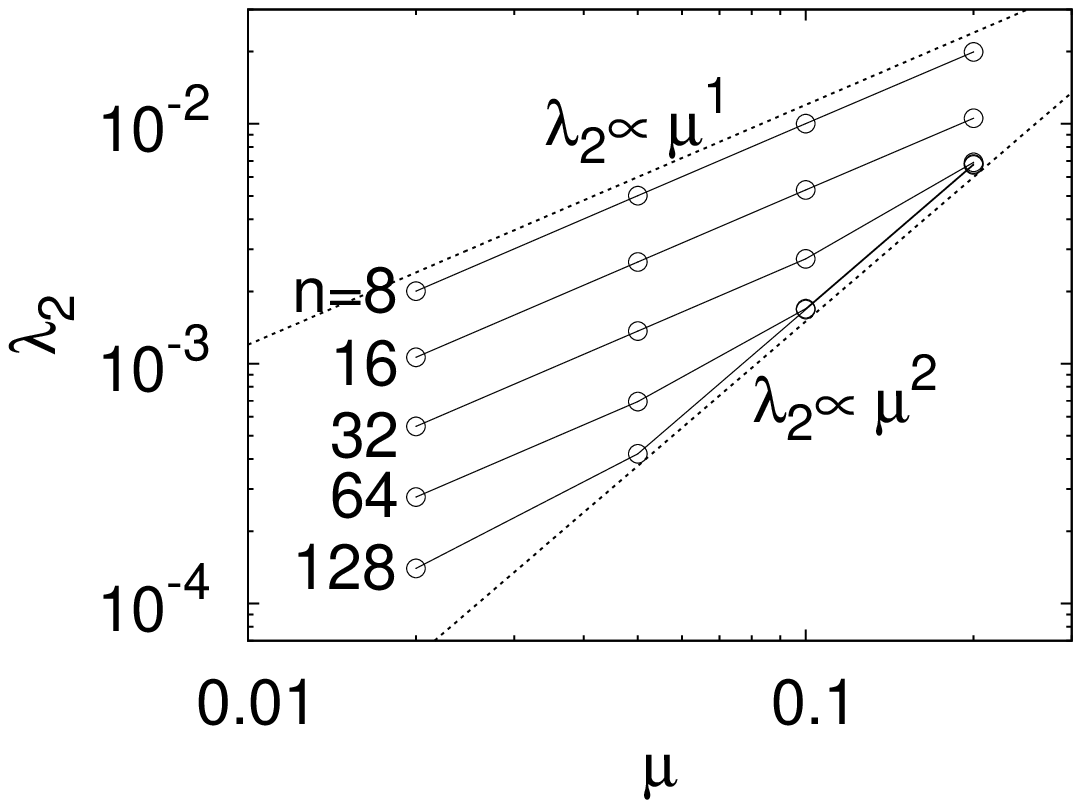}\includegraphics[width=1.7in]{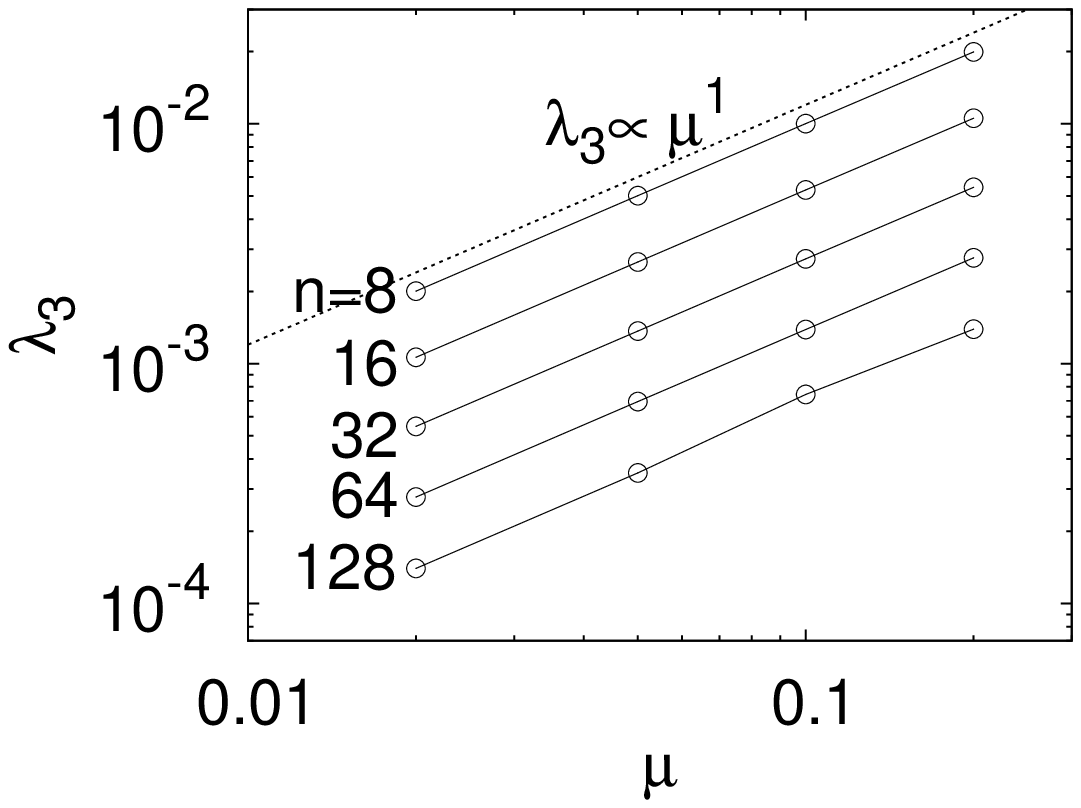}}
\centerline{\includegraphics[width=1.7in]{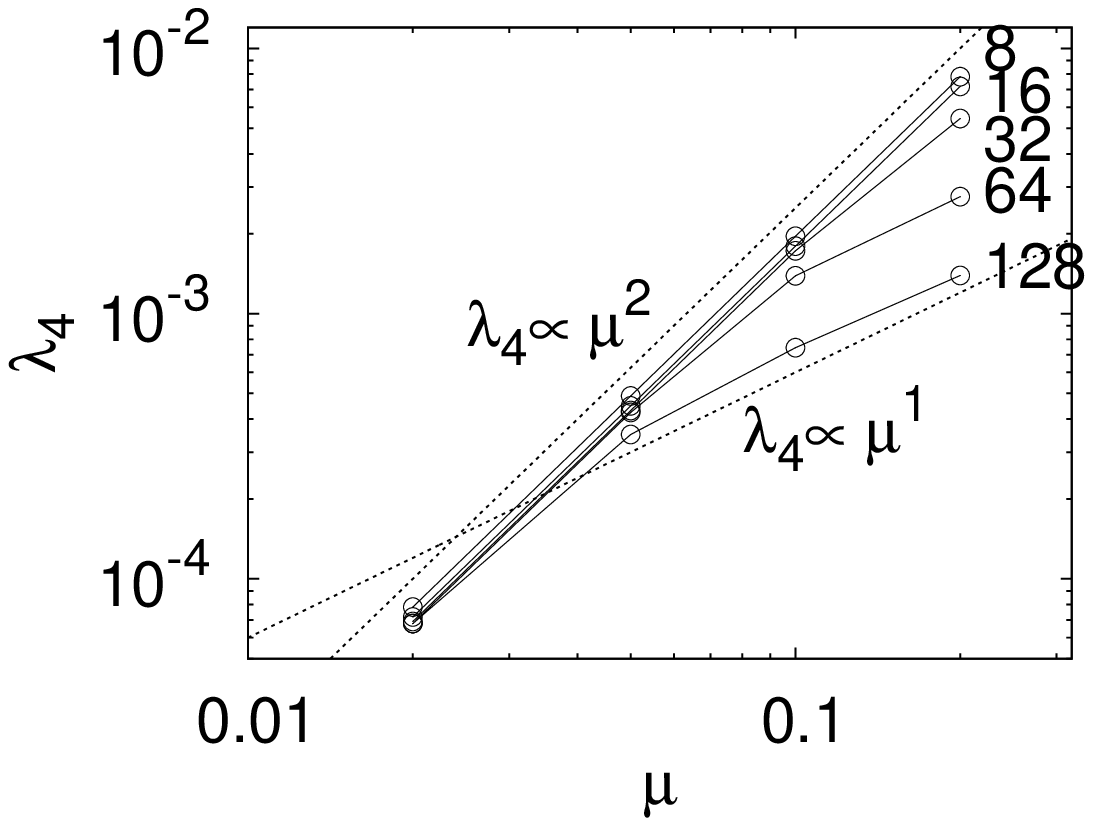}\includegraphics[width=1.7in]{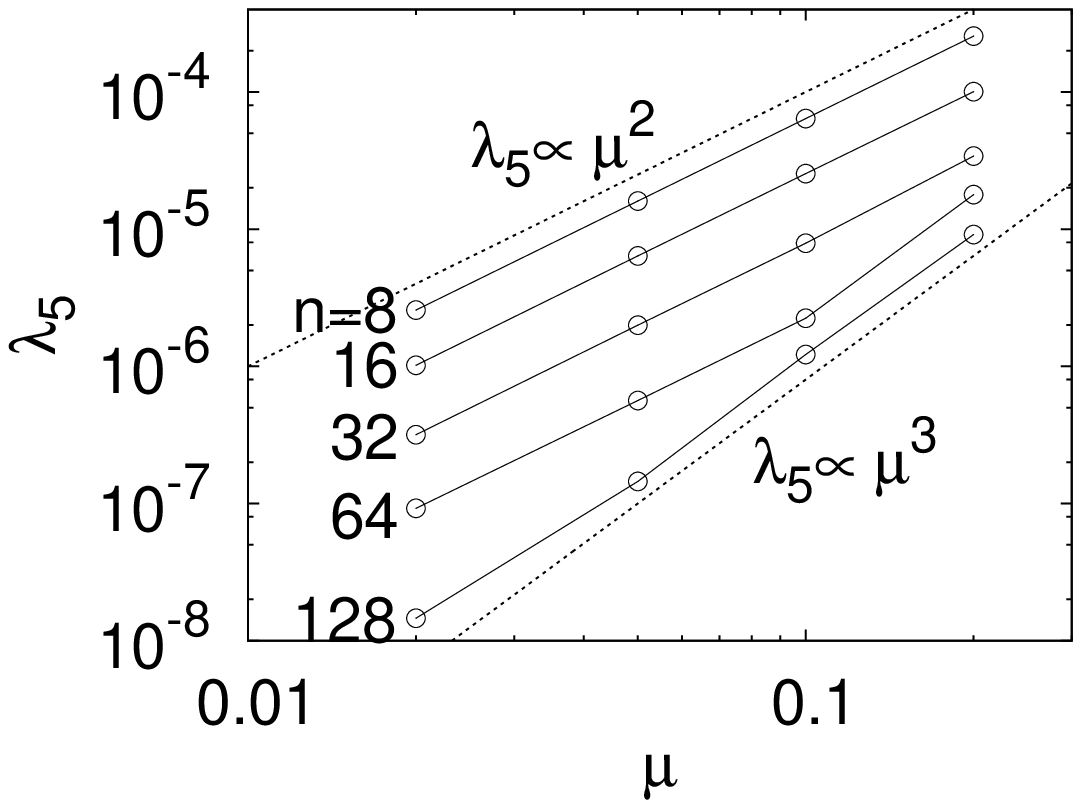}}
\caption{Scaling of the Schmidt coefficients $\lambda_{2,3,4,5}$ with $\mu$, obtained by the tDMRG (circles) for $n=8,16,32,64,128$. Dashed lines suggest the asymptotic scaling with $\mu$. We use a symmetric bipartite cut after $n/2$ spins, $\Gl=\Gr=\gamma=1$, $\bar{\mu}=0$.}
\label{fig:skal}
\end{figure}
Looking at Fig.~\ref{fig:skal}, and noting that the transition point of the avoided crossing $\mu\, n = \nu_{\rm c}$ happens at $\mu \approx 0.05$ for $n=128$ and $\mu \approx 0.1$ for $n=64$, we can observe the following scaling for the largest coefficients: (i) $\lambda_2 \sim \mu$ for $\mu\, n < \nu_{\rm c}$, while $\lambda_2 \sim \mu^2$ for $\mu\, n > \nu_{\rm c}$; (ii) $\lambda_3 \sim \mu$; (iii) $\lambda_4 \sim \mu^2$ for $\mu\, n < \nu_{\rm c}$, while $\lambda_4 \sim \mu$ for $\mu\, n > \nu_{\rm c}$; (iv) $\lambda_5 \sim \mu^2$ for $\mu\, n < \nu_{\rm c}$, while $\lambda_5 \sim \mu^3$ for $\mu\, n > \nu_{\rm c}$. We can see that for $\mu\, n > \nu_{\rm c}$ $\lambda_5$ decays faster with $\mu$ than $\lambda_4$. The MPO with $D=4$ (\ref{eq:MPO}) therefore also gives the leading order solution in the limit of weak driving, $\mu \to 0$, provided we have $\mu\, n > \nu_{\rm c}$. Presumably there are further avoided crossings in the spectrum of Schmidt coefficients, besides 
 the one at $\mu\, n = \nu_{\rm c}$. $\mu\, n$ should therefore be smaller than the value at these higher crossings, however, these points need further investigation.

To summarize, an MPO ansatz of size $D=4$ (\ref{eq:MPO}) gives the leading order solution in two limits, either $n \to \infty$ while keeping $\mu\, n < \nu_{\rm c}$, or $\mu \to 0$ while keeping $\mu\, n > \nu_{\rm c}$. The two conditions can in fact be put under the same hood. Observing that to fulfill $n\,\mu < \nu_{\rm c}$ in the thermodynamic limit one must necessarily have $\mu \to 0$, and similarly, to have $n\,\mu > \nu_{\rm c}$ in the weak-driving limit one must necessarily have $n \to \infty$, one can reformulate both in a single statement, saying that the MPO of dimension $D=4$ (\ref{eq:MPO}) gives the leading order solution in the limit $n \to \infty$ having at the same time $\mu \to 0$.

Another interesting point is that for $\mu\, n > \nu_{\rm c}$ the second largest Schmidt coefficient is independent of $n$ and scales as $\lambda_2 \sim \mu^2/n^0$. This means that despite the fact that our system is at an infinite effective temperature, there is a nonzero bipartite entanglement present, even in the thermodynamic limit of large $n$ (at a fixed $\mu$). This entanglement at an infinite temperature is of purely nonequilibrium origin.

\section{Nonequilibrium phase transition}

From the exact solution we can see that the NESS undergoes a transition from a state without long-range correlations for $\gamma=0$, to the one with long-range correlations for $\gamma \neq 0$. A point in a parameter space where a system undergoes a sudden change in some expectation values is usually called a phase transition point. We use the same nomenclature here and call this transition a nonequilibrium phase transition, because the nature of the correlations changes. Transport properties also change suddenly at $\gamma=0$, going from a ballistic (superconducting) state to a diffusive for nonzero dephasing. In the phase with long-range correlations two-point $z-z$ correlations scale as $\sim \mu^2/n$~\cite{JSTAT10} and are therefore of purely nonequilibrium origin. They also go to zero in the thermodynamic limit making this transition different than equilibrium phase transitions. The correlation function has a plateau because in the thermodynamic limit the decay of $C_{i,
 j}$ with the distance between indices $|i-j|$ gets increasingly slower (\ref{eq:plat}). Recently, similar quantum nonequilibrium phase transitions have been discovered in the XY model~\cite{Iztok08,bojan}, in which the long-range correlations scale as $\sim \mu^2/n$, in the XXZ model with dephasing~\cite{NJP10} (the same scaling of the correlation plateau as here), as well as in the XXZ model without dephasing~\cite{PRL10}, where though the correlation plateau scales as $\sim \mu^2/n^0$ (the plateau is independent of $n$!) at an infinite temperature and as $\sim \mu^2/n^\alpha$ at a finite temperature. It has been conjectured that long-range correlations are a generic feature of quantum nonequilibrium steady states~\cite{PRL10}.

It would be nice to understand this phase transition more in detail. At equilibrium, phase transitions are connected with the nonanaliticity of the free energy, or equivalently, zeros of the partition function. Difficulty with the nonequilibrium situation is that there is no general theory, in particual, free-energy-like quantity whose analytic property could be studied is not known. Nevertheless, because observable properties of the system change suddenly at our nonequilibrium phase transition point there must be some underlying nonanalytic property. We can mention that the Lee-Yang theory of equilibrium phase transitions has been used with some success on certain classical nonequilibrium systems~\cite{arndt,blythe,Evans:02} whose stationary state can be represented in a matrix product form. For some classical exclusion processes in the thermodynamic limit even a free-energy like functional can be calculated~\cite{Derrida:01}. Until recently~\cite{JPA10} no quantum stationar
 y state solvable by matrix product ansatz has been known and the theory of quantum nonequilibrium phase transitions is lagging behind the classical. For classical nonequilibrium systems a product of all nonzero eigenvalues of a superoperator can play a role of a ``partition'' function, $Z=\prod_{\lambda_j \neq 0}(-\lambda_j)$, whose zeros then determine the location of nonequilibrium phase transitions. It is shown~\cite{Evans:02} that the product of nonzero eigenvalues is equal to a sum of all expansion coefficients of the NESS. There are problems with such an approach though. The principal difficulty is that $Z$ is defined only up-to to an arbitrary normalization factor and it is a priori not known which normalization one should take.

\begin{figure}[!h]
\centerline{\includegraphics[width=3.5in]{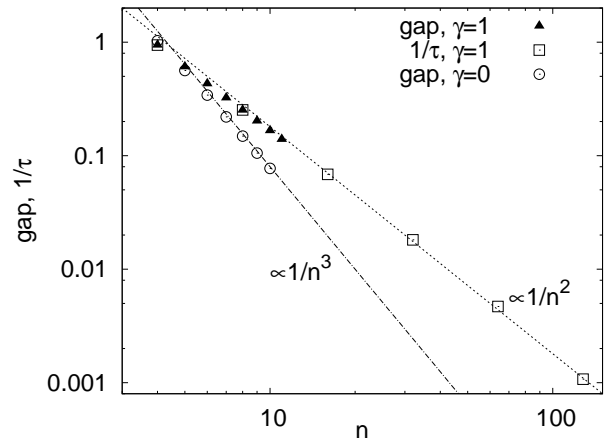}}
\caption{Scaling of the gap of the Lindblad superoperator with system size for the model without long-range order at $\gamma=0$ and for the system with long-range order in the case of $\gamma=1$. At the nonequilibrium phase transition point at $\gamma=0$ the gap decays faster, as $\Delta \sim 1/n^3$, than away from the transition point ($\gamma=1$) where $\Delta \sim 1/n^2$. We also show the scaling of the inverse relaxation time of the magnetization at site $n/2$ (squares). $\Gl=\Gr=1$, $\bar{\mu}=0$ and $\mu=0.1$.}
\label{fig:gap}
\end{figure}

In the present work we look for non-analytic signatures in the spectrum of the superoperator, trying to see if there is any characteristic change at the phase transition point. Because the NESS is an eigenstate of ${\cal L}$ with an eigenvalue zero, a change in the properties of the NESS should be connected with the closing of the gap between the second largest eigenvalue and the one of the NESS. If this gap goes to zero there is the possibility for a scenario reminiscent of an avoided-crossing between two eigenenergies happening at a zero-temperature quantum phase transition, where at the crossing the nature of the two levels involved is exchanged. Of interest for nonequilibrium phase transitions is therefore the second largest eigenvalue $\lambda_2$ of the Lindblad superoperator ${\cal L}$, or in particular the gap $\Delta$, $\Delta =-\lambda_2$. We note that the size of the gap also determines the decay rate, i.e., the relaxation rate, to the NESS. Namely, the initial nons
 tationary state will, for large times, relax to the NESS as $\exp{(-\Delta\, t)}$. Because we in general expect the relaxation time, in the case of local coupling to baths at chain ends, to grow with the system size at least as $\propto n$ -- larger system simply needs more time to relax --, the gap is expected to decrease with $n$, even if we are not at the nonequilibrium phase transition point. The signature of nonequilibrium phase transition can therefore not be simply the closing of the gap $\Delta$. What is generally observed though is that the closing of the gap $\Delta$ with $n$ is at the phase transition point {\em faster} than away from the phase transition point. Such behavior has been observed, for instance, in an open XY chain~\cite{Iztok08}, where at a nonequilibrium phase transition point the gap scales as $\Delta \sim 1/n^5$, while it scales only as $\Delta \sim 1/n^3$ away from the transition point. We therefore studied the scaling of the gap in our model. Th
 e results can be seen in Fig.~\ref{fig:gap}.
We numerically exactly calculated the gap $\Delta$ for systems of up-to $n=11$ spins. For larger systems a relaxation time $\tau$ of some observable can be used as an estimate of the gap. Using the tDMRG we have solved the Lindblad equation in time, thereby obtaining time dependent expectations of observables. We fitted an exponential function to the relaxation of the magnetization at site $n/2$, $z_{n/2}(t)-z_{n/2}(\infty)\sim \exp{(-t/\tau)}$, and determined $\tau$ for systems with up-to $n=128$ spins. As the second largest eigenvalue $\lambda_2$ is nondegenerate for $\gamma=1$ the relaxation time can serve to estimate the gap through $\Delta \sim 1/\tau$. From the figure we can see that at the phase transition point the gap scales as $\Delta \sim 1/n^3$. This of course agrees with a previous analytic result for the isotropic XY chain~\cite{Iztok08}. Away from the nonequilibrium phase transition, at $\gamma=1$, the gap decays more slowly, as $\Delta \sim 1/n^2$. We therefor
 e conjecture that the characteristic feature of nonequilibrium phase transitions is that the gap of the superoperator decays faster at the nonequilibrium phase transition point than in its neighborhood. At a nonequilibrium phase transition point the relaxation slows down, similary as at an equilibrium phase transition point. How much faster the gap decays depends on a particular system. Here we have $1/n^3$ vs. $1/n^2$, for the XY model on the other hand one has $1/n^5$ vs. $1/n^3$~\cite{Iztok08}. Note that this criterion is different from those used in equilibrium physics as it makes a reference also to the neighborhood of the transition point. Interesting to note is that the relaxation time of the spin current is in our model by about a factor $\approx 4$ shorter than that of $z_{n/2}$. The reason must lie in the fact that some of the eigenstates that are just below the NESS in the spectrum carry no current. Another observation about the spectrum of the Lindblad superopera
 tor for the XX model with dephasing is that the eigenvalues are independent of the driving $\mu$, only the eigenvectors depend on $\mu$. This probably happens due to the hierarchical structure of the master equation.

\section{Summary}

We have provided exact expressions for all one-point, two-point and three-point connected correlations in the nonequilibrium stationary state of the XX model with dephasing. A hierarchical structure of stationary equations is explained, a consequence of which is that equations determining all $n$-point correlations form a closed set. The nonequilibrium stationary state is non-Gaussian because the Wick theorem does not apply. In the thermodynamic and weak-driving limit the solution can be written in terms of a matrix product ansatz with matrices of fixed dimension 4. At zero dephasing the model exhibits a nonequilibrium phase transition from a state with only nearest-neighbor correlations to a state possessing long-range correlations. It is conjectured that at a quantum nonequilibrium phase transition point the gap of the superoperator closes with the system size more rapidly than in the vicinity of the transition point.

\begin{acknowledgements}
Support by the Program P1-0044 and the Grant J1-2208 of the Slovenian Research Agency, the project 57334 by CONACyT, Mexico, project IN114310 by the University of Mexico, as well as discussion with T.~Prosen, is acknowledged.
\end{acknowledgements}

\appendix*
\section{3rd order terms}
\label{app:A}
We are mainly interested in the connected three-point correlations, defined for arbitrary operators $A,B,C$ as $\ave{ABC}_{\rm c}=\ave{ABC}-\ave{A}_{\rm c} \ave{B}_{\rm c}\ave{C}_{\rm c}-\ave{AB}_{\rm c}\ave{C}_{\rm c}-\ave{AC}_{\rm c}\ave{B}_{\rm c}-\ave{BC}_{\rm c}\ave{A}_{\rm c}$. The 3rd order ansatz $\mu^3 R^{(3)}$ is the sum of 5 terms, 
\begin{equation}
\mu^3 R^{(3)}=G_{\rm zzz}+G_{\rm zzj}+G_{\rm zjj}+G_{\rm jjj}+G'_{\rm jjj}.
\end{equation}
The terms are
\begin{widetext}
\begin{eqnarray}
G_{\rm zzz} &=& \sum_{i=1}^n \sum_{j=i+1}^n \sum_{k=j+1}^n (Z_{i,j,k}+a_i a_j a_k+a_i C_{j,k}+a_j C_{i,k}+a_k C_{i,j}) \sigma_{\rm i}^{\rm z} \sigma_{\rm j}^{\rm z}\sigma_{\rm k}^{\rm z}\nonumber \\
G_{\rm zzj} &=& \sum_{k=1}^{n-1} \sum_{i=1 \atop i \notin \{k,k+1 \}}^n \sum_{j=i+1 \atop j \notin \{k,k+1\}}^n \frac{1}{2}(X_{k,i,j}+a_i a_j b+b C_{i,j}+a_i\ave{j_k \sigma^{\rm z}_j}_{\rm c} + a_j\ave{j_k \sigma^{\rm z}_i}_{\rm c}) j_k \sigma_{\rm i}^{\rm z}\sigma_{\rm j}^{\rm z} \nonumber \\
G_{\rm zjj} &=&  \sum_{i,k=1 \atop k>i}^{n-1}\sum_{j=1 \atop j \notin \{i,i+1,k,k+1 \}}^n (Y_{j,i,k}+a_j b^2+a_j(f-b^2)+b\ave{\sigma_j^{\rm z}j_k}_{\rm c}+b\ave{\sigma_j^{\rm z}j_i}_{\rm c}) \sigma_j^{\rm z} \frac{j_i j_k+j_k j_i}{8} \nonumber \\
G_{\rm jjj} &=& \sum_{i,k,l=1 \atop i\neq k,i\neq l,k\neq l; \hbox{at most one overlap}}^{n-1} \frac{w}{8} j_i j_k j_l \nonumber \\
G'_{\rm jjj} &=& \sum_{i=1}^{n-3} g\, (\sigma_i^{\rm x}\sigma_{i+3}^{\rm y}-\sigma_i^{\rm y}\sigma_{i+3}^{\rm x}) .
\end{eqnarray}
\end{widetext}
In the above expressions the summations are such that no two operators appear at the same site and in addition there are no overlapping terms like $j_k \sigma^{\rm z}_{k+1}$. In the $G_{\rm jjj}$ we have at most one overlapping currents on neighboring sites, like $j_k j_{k+1}$. The solutions for unknown coefficients are
\begin{equation}
w=b^3\frac{(1+t)^2}{6t(t-1)},\qquad g=b^3\frac{(1+t)^2}{t(t-1)},
\end{equation}
\begin{widetext}
\begin{equation}
Z_{i,j,l}= -b^3\frac{2}{t(t-1)} \left\{ k^{(\rm L)}_i(k^{(\rm R)}_{j}-k^{(\rm L)}_j)k^{(\rm R)}_{l}+(t+1)(k^{(\rm R)}_{l} \delta_{j,i+1}-k^{(\rm L)}_i \delta_{j,l-1}) \right\}.
\label{eq:zzz}
\end{equation}
\begin{equation}
X_{k,i,j}=-b^3\frac{2}{t(t-1)} \cdot \left\{
\begin{array}{lcl}
 k^{(\rm L)}_i(k^{(\rm R)}_{j}-k^{(\rm L)}_j)+(t+1)\delta_{j,i+1}  &;& \hbox{$k>j$} \\
-k^{(\rm R)}_{j}(k^{(\rm R)}_{i}-k^{(\rm L)}_i)+(t+1)\delta_{j,i+1}  &;& \hbox{$k<i$} \\
 2k^{(\rm L)}_i k^{(\rm R)}_{j} &;& \hbox{$i<k<j-1$}\\
\end{array}
\right. .
\label{eq:3rdX}
\end{equation}
\end{widetext}
\begin{equation}
Y_{j,i,k}=-b^3\frac{2}{t(t-1)} \cdot \left\{
\begin{array}{lcl}
-2k^{(\rm R)}_{j}  &;& \hbox{$j>k+1$} \\
 2k^{(\rm L)}_j &;& \hbox{$j<i$} \\
 (k^{(\rm L)}_j- k^{(\rm R)}_{j}) &;& \hbox{$i+1<j<k$}
\end{array}
\right. .
\label{eq:3rdY}
\end{equation}
Three-point connected correlation function is for nondiagonal indices equal to $\ave{\sigma_i^{\rm z} \sigma_j^{\rm z} \sigma_k^{\rm z}}_{\rm c}=Z_{i,j,k}$.

\end{document}